\documentclass[preprint,12pt,authoryear]{elsarticle}

\usepackage{amssymb}
\usepackage{graphicx}
\usepackage{dcolumn} 
\usepackage{bm} 
\usepackage{color}

\journal{Physica D}

\begin{document}

\begin{frontmatter}



\title{Stability and onset of two-dimensional viscous fingering in immiscible fluids}
\author{Radha Ramachandran}


\address{The James Franck Institute and Department of Physics, The University of Chicago, Chicago, IL 60637}

\begin{abstract}
Viscous flows in a quasi-two-dimensional Hele-Shaw geometry can lead to an interfacial instability when one fluid, of viscosity $\eta_{in}$ displaces another of higher viscosity, $\eta_{out}$.  Recent studies have shown that there is a delay in the onset of fingering in miscible fluids as the viscosity ratio, $\eta_{in}/\eta_{out}$, increases and approaches unity; the interface can remain stable even though the displacing liquid is less viscous. This paper shows that a delayed onset and stable pattern can occur in immiscible fluids as well.  However, there are two significant differences between the two cases. First, in miscible fluids, stable patterns are obtained whenever $\eta_{in}/\eta_{out} > 0.33$ while in immiscible fluids, the radius at which the onset of fingering starts, $R_{onset}$, increases steadily until $\eta_{in}/\eta_{out}=1$.   A stable pattern is obtained only when the total size of the plate used is smaller than $R_{onset}$. Second, once the delayed fingering starts in immiscible fluids, the fingers grow faster than the central circular region. In miscible fluids, there is a regime in which the fingers and the central region grow in proportion to each other. These differences between miscible and immiscible fingering are maintained even when we compare immiscible fluids that have very low interfacial tension with miscible fluids that have negligible diffusion. 

\end{abstract}

\begin{keyword}
Viscous fingering \sep radial Hele-Shaw \sep immiscible \sep fingering onset \sep stable patterns \sep 3D tongue shape

\PACS 47.15.gp \sep 47.20.Gv \sep  47.54.-r \sep 47.55.N-


\end{keyword}

\end{frontmatter}



\section{Introduction and background}
When a fluid of viscosity $\eta_{in}$ displaces a second fluid of viscosity $\eta_{out}$ confined within a thin gap, the interface between the two fluids can become unstable forming finger-like protrusions. Such a finger will grow and can repeatedly split leading to the formation of complex viscous-fingering patterns (\cite{paterson1981radial}). This instability is often studied in the quasi two-dimensional geometry of a Hele-Shaw cell which consists of two thick glass plates with a small spacing between them (\cite{chen1987radial, nittman1985fractal, daccord1986radial, bensimon1986viscous, homsy1987viscous, paterson1981radial, tabeling1987experimental, mccloud1995experimental, miranda1998radial, goyal2006miscible, moore2002fluctuations}). Understanding the viscous-fingering instability has applications in industrial processes such as in petroleum extraction, carbon sequestration or sugar refining (\cite{hill1952channeling, orr1984use, white2012co2, wang2004unstable, cinar2007experimental, araktingi1993viscous, weitz1987dynamic}) as similar fingering patterns are also observed in porous medium flows. 

The seminal work by\cite{saffman1958penetration}, showed that the interface between the two liquids becomes unstable in this quasi-two-dimensional geometry whenever the \emph{displacing} liquid is less viscous than the \emph{displaced} liquid. They showed that most unstable wavelength, $\lambda_c$, that determines the typical finger width, depends on the difference in the fluid viscosities,  $\Delta \eta \equiv \eta_{out} - \eta_{in}$, as well as the interfacial tension between the fluids, $\sigma$, the interfacial velocity, $V$, and the plate spacing, $b$:
\begin{equation}
\lambda_c = \pi b\sqrt{\frac{\sigma}{V \Delta \eta}}.
\label{saff-t}
\end{equation}

In the case of miscible fluids, the role played by diffusion in determining $\lambda_c$ has been analyzed in detail \cite(de2005viscous, Mishra2008, graf2002density, chen2001miscible, jha2011fluid, tan1988simulation, perkins1965mechanics)
However, when the plate spacing, $b$, becomes comparable to the unstable wavelength, $\lambda_c$, the typical finger width is independent of $V$, $\sigma$ and $\Delta \eta$ and is determined instead only by the plate spacing (\cite{paterson1985fingering}):
\begin{equation}
\lambda_c \propto b
\label{patt}
\end{equation}
~A large literature has been dedicated to measuring and understanding the control parameters that determine the fingering patterns in different fluid systems (\cite{homsy1987viscous, tabeling1987experimental, mccloud1995experimental, miranda1998radial, goyal2006miscible, moore2002fluctuations, lindner2002viscous, cheng2008towards, pihler2012suppression, li2009control}). 

Many viscous fingering experiments have been performed in the limit of $\eta_{in}/\eta_{out} \ll1$ (\cite{chen1987radial, saffman1958penetration, paterson1981radial, FLM:392852, praud2005fractal, PhysRevE.70.016303}).  Recently, another prominent feature was discovered in a radial Hele-Shaw cell when the viscosity ratio, $\eta_{in}/\eta_{out}$, was varied by about three orders of magnitude for both miscible and immiscible fluids (\cite{irmgard, radar}). In these experiments, the fluids were injected through a hole near the center of the glass plates so that the interface expands radially outward. The pattern has a central circular region, where the outer fluid is completely displaced and  fingers appear only beyond the radius of this circle. 

While $\lambda_c$, which depends on $\Delta \eta$, dictates how \emph{wide} the fingers are, it is the viscosity ratio, $\eta_{in}/\eta_{out}$, that determines how \emph{long} the fingers will be with respect to the central stable region; as $\eta_{in}/\eta_{out}$ decreases, the finger length increases with respect to the central stable region. Hence, varying $\eta_{in}/\eta_{out}$ allows a study of the transition from completely stable circular patterns to fingering patterns. Such a study for miscible fluids with negligible diffusion, where the Peclet number, $Pe= bV/D > 2000$ (where $D$ is the mass diffusion coefficient), showed two peculiar behaviors in the circular geometry~(\cite{irmgard}): (i) as also found in a linear geometry (\cite{lajeunesse19973d}), and unlike predictions by Saffman and Taylor, stable patterns can occur even when $\eta_{in}/\eta_{out}<1$; (ii) for a range of $\eta_{in}/\eta_{out}$, there is a delay in the onset of fingering. That is, fingers are formed at a later time, only after an initial circular region is formed. Even after the fingers form, they do not grow rapidly and show what is known as ``proportionate'' growth, i.e., fingers grow in proportion to the inner circle and do not split or form branches (\cite{irmgard, radar}). This kind of growth though common in mammalian growth, is rarely seen in physical systems (\cite{sadhu2012modelling, dhar2013sandpile}). 

The present paper studies the transition from stable to unstable pattern formation in immiscible fluids in a circular Hele-Shaw geometry.  It considers two cases, one with high interfacial tension (24 to 29 mN/m) and another with low interfacial tension (1 to 1.2 mN/m) in order to understand the role played by interfacial tension and to examine whether the limit of low interfacial tension in immiscible fluids approaches the behavior of miscible fluids.  Similar to the case of miscible fluids when $\eta_{in}/\eta_{out} < 1$, there is a delayed onset of fingering where the radius of the displacing fluid must grow to a certain radius, $R_{onset}$, before fingers start to form. This onset radius increases as $\eta_{in}/\eta_{out} \rightarrow 1$.  This delay is larger than what is predicted due to the stabilizing effects of circular geometry of the plates (\cite{al2012control}). Seemingly stable patterns can form when $R_{onset}$ becomes comparable to the size of the plates used. This is in contrast to miscible fluids, where the plate size has no effect on whether the pattern becomes stable or unstable.  There is no regime of proportionate growth.
 
Immiscible fluids have similar finger lengths irrespective of whether the interfacial tension is low or high;  these are \emph{longer} fingers than those found in miscible fluids with negligible diffusion.   Hence, one must look for another explanation to understand why miscible fluids have shorter fingers.  By looking at the cross-sectional profile of the displacing fluid one can look for three-dimensional (3D) effects in suppressing finger formation. There is a significant difference in the 3D shape of the displacing front between the miscible and immiscible cases. Even the small amount of interfacial tension in immiscible fluids is enough to produce a \emph{flat} displacing front, whereas miscible fluids have a protruding tongue in between the two plates at the displacment front. This difference could contribute to the slowing down of finger formation in miscible fluids. 
 
\section{Experimental Methods}
\label{methods}

The experiments are conducted in a radial Hele-Shaw cell, using two sets of 1.9 cm thick, circular, glass plates of radii 14 cm and 25 cm. These plates are maintained at constant gap using spacers of 102 $\mu$m to 635 $\mu$m thickness. Liquids are injected through a 1.6 mm hole in the center of the plates using a syringe pump (New Era Pump Systems NE1010). 

To ensure that the initial interface between the outer and inner fluid is uniform, we drill holes in the center of both the top and bottom  plates. After the more viscous fluid has completely filled the gap between the two plates, the less viscous fluid is injected through the hole in the top plate and then allowed to flow out through the hole in the bottom plate.  This provides an initial cylindrical column of the fluid that completely spans the gap between the plates. When the hole in the bottom plate is closed, this cylindrical column of fluid expands radially between the plates. 

For the immiscible case, two pairs of fluids with different interfacial tensions, $\sigma$, were used: (a) Silicone oils (Clearco) and mixtures of glycerin-water (Fischer Scientific) with interfacial tensions ranging between 24 mN/m to 29 mN/m and (b) Silicone oils and mineral oils (Fischer Scientific) with interfacial tensions ranging between 1 to 1.2 mN/m. Inner fluids are dyed using Brilliant Blue G dye (Alfa Aesar) or Oil red O (Sigma-Alrich). The results are compared with earlier experiments done with miscible fluids \cite{irmgard}.   While it is known that diffusion can play an important role in stabilizing an interface (\cite{erglis2013magnetic, kitenbergs2015magnetic}), the experiments reported here have used fluids with high viscosities to minimize any diffusion. These pairs of miscible fluids have a high Peclet number, so that the flows in the experiments are dominated by advection and there is negligible diffusion at the interface.

The interfacial tensions for the immiscible fluids are measured using the pendant drop method (\cite{ARASHIRO1999}). Here a drop of one fluid is gravitationally suspended inside a less dense one. The drop is stationary due to the balance between gravitational and interfacial-tension forces. Because we know the gravitational force acting on the fluid, we can analytically solve for the shape of the drop to obtain the interfacial tension.  The values obtained using this method agree with literature values (\cite{than1988measurement}).  Fluid viscosities are measured using calibrated viscometers (Cannon-Ubbelohde). The fingering patterns are recorded using a Prosilica GX camera at 2 to 15 frames/s. 

\begin{figure}
\centering 
\begin{center} 
\includegraphics[width = 0.4\columnwidth]{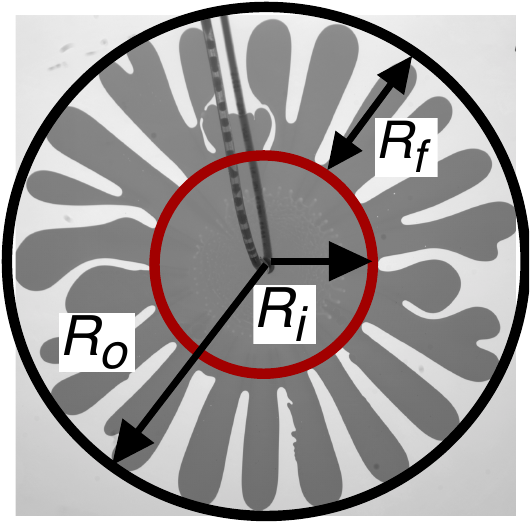} 
\caption{ Definition of length-scales to characterize the structure of the viscous fingering patterns. $R_i$ is the largest circle that can be inscribed completely inside the inner fluid. $R_o$ is the smallest circle enclosing the entire pattern. $R_f \equiv R_o - R_i$. Image taken from Ref. \cite{radar}}
\label{size-ratio}
\end{center}
\end{figure}

To characterize the overall structure of the patterns, three length-scales are measured, as shown in Fig. \ref{size-ratio} taken from \cite{radar}. The inner radius, $R_i$, is defined as the largest circle completely inscribed in the inner fluid. The outer radius, $R_o$, is the smallest circle completely enclosing the inner fluid.

To measure $R_i$ and $R_o$, the images are binarized so that there is a black pixel wherever there is inner fluid and a white pixel wherever there is outer fluid. (This ignores any variation in the intensity of the dye within the inner fluid, that is caused by the inner fluid not completely displacing the outer fluid across the gap).  A very small circle is initially placed inside the inner fluid and expanded it until it touches the interface. The center of the circle is then moved away from the interface in the direction given by a radial line joining the center of the circle to the point of intersection.  When this is done, the circle no longer intersects the interface. This process of iteratively expanding the circle and moving the center away from the intersection is continued until the circle intersects the interface at multiple points inside the pattern so that it can no longer move. The radius of this inner circle is $R_i$.  Similarly, to obtain $R_o$, the largest circle that can be placed in the image is continually decreased until it intersects the interface of the pattern. The circle is then iteratively moved away from the interface until it no longer touches it; the size of the circle is then reduced. This process is continued until the circle touches multiple points on the interface so that it can no longer be moved. This outer circle has radius $R_o$. 

The length of the finger, $R_f$, is then given by  $R_o - R_i$. The size-ratio, {defined as} $R_f/R_i$, increases rapidly at the beginning of injection, but grows very slowly at late times as the pattern increases in size.  To capture the dynamics at long times, we measure $R_f/R_i$ at the point where the outer radius, $R_o$, is 8 cm. 

It has been shown previously that the inner fluid forms a tongue at the center between the two plates (\cite{FLM:392852,lajeunesse19973d,irmgard, setu2013viscous, yang1997asymptotic}). The thickness of these tongues is measured by using the absorption of light due to the dye in the inner fluid. Here, the measurement is calibrated using a wedge of known opening angle that is filled with the inner fluid with a known amount of dye. On keeping the incident light constant across the wedge, the light intensity measured through the wedge depends on the thickness of the fluid layer. This produces a calibration curve that allows intensity values to be converted into a three-dimensional finger profile.  

The delay in the onset of fingers, is characterized by the onset radius, $R_{onset}$, defined as the average distance from the center at which the interfacial instability can first be observed.  

\section{Immiscible vs miscible displacements: a comparison of finger lengths}

\begin{figure}
\centering 
\begin{center} 
\includegraphics[width = 0.85\columnwidth]{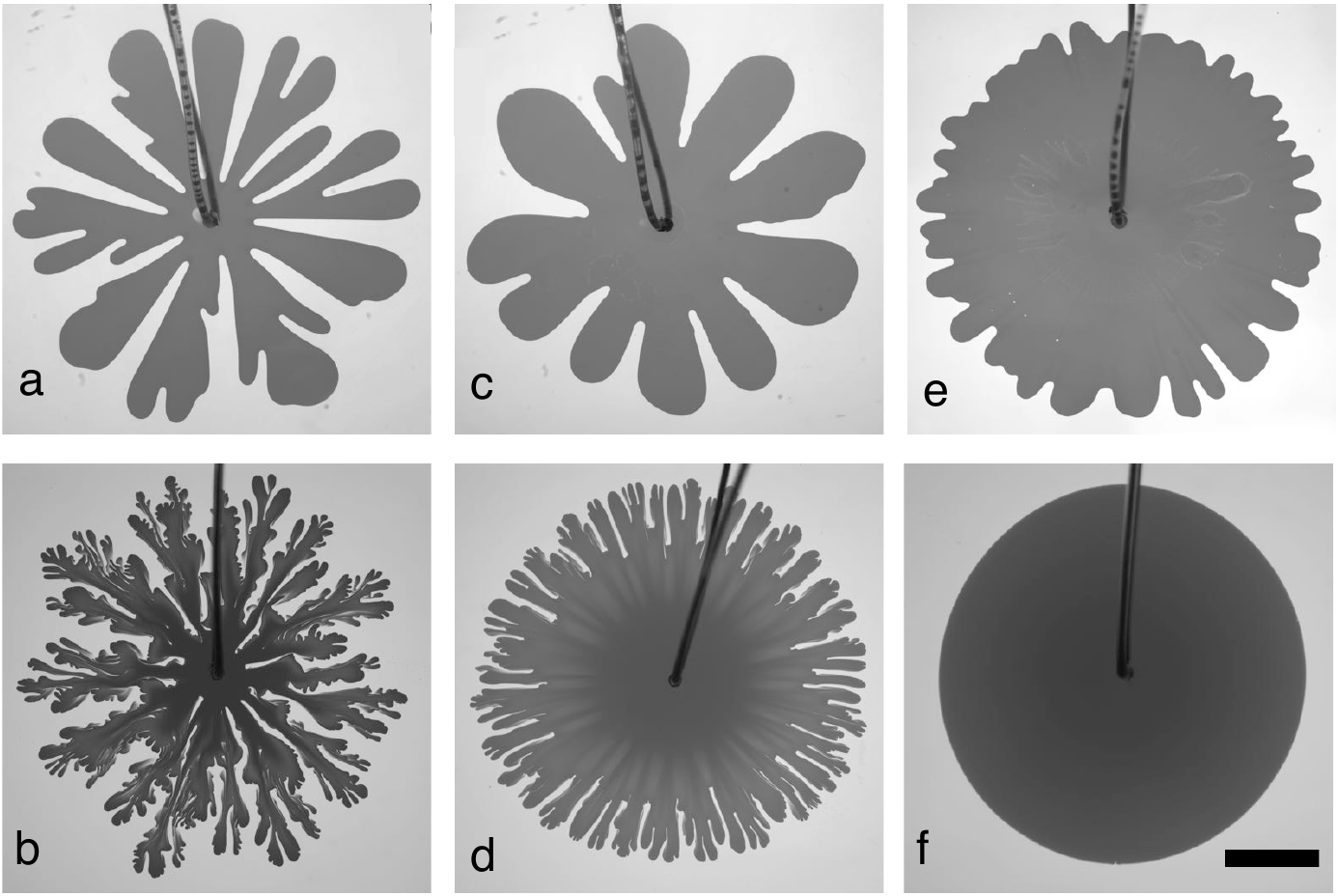} 
\caption{ A comparison between immiscible (top) and miscible (bottom) fingering patterns at different viscosity ratios. (a), (c), (e) show patterns formed when silicone oil is displaced by glycerol-water mixtures; the finite interfacial tension leads to wider fingers.  (b), (d), (f) show glycerin displaced by glycerin-water mixtures of lower viscosities, (a and b) $\eta_{in}/\eta_{out}$ = 0.0064; (c) $\eta_{in}/\eta_{out}$ = 0.067 and (d) $\eta_{in}/\eta_{out}$ = 0.068; (e and f) $\eta_{in}/\eta_{out}$ = 0.38. At the two larger values of $\eta_{in}/\eta_{out}$, it is clear that the immiscible fluid displacements produce longer fingers compared to those found in the miscible fluids. Scale bar is 4 cm.}
\label{misc-immisc}
\end{center}
\end{figure}

A visual comparison of the fingering patterns seen in immiscible and miscible fluids is shown in Figure \ref{misc-immisc}. The images in the top row show immiscible fluid displacements and in the bottom row show miscible displacements. In each column, $\eta_{in}/\eta_{out}$ are adjusted to be nearly the same for both the immiscible and miscible fluid pairs.  For both immiscible and miscible fluids, as one goes from left to right, $\eta_{in}/\eta_{out}$ increases and the finger length decreases with respect to the radius of the central stable region. 

In the first column, the finger lengths of miscible and immiscible fluids are comparable, whereas the \emph{width} of the fingers are greater for immiscible fluids as it has higher $\sigma$ (and therefore higher $\lambda_c$).  In the second column, the immiscible pair of fluids has slightly longer as well as wider fingers than the miscible pair.  A noticible difference here is that, while the interior region of the immiscible fluids appears uniformly dark, the interior region of miscible fluids shows a gradation of color, pointing to an incomplete displacement of the outer by the inner fluid. To ensure that the gradation of color noticed is indeed incomplete displacement, and not diffusion, a UV-curable polymeric fluid was used as one of the miscible fluids and was ``frozen''  by UV light exposure (\cite{irmgard}). In the third column, immiscible fluid displacements are unstable whereas the miscible fluid displacements are stable. An analysis of Fig. \ref{misc-immisc}f, shows that there continues to be a gradation of color pointing to an incomplete displacement of outer fluid. This is discussed in detail in Section \ref{tongue-section}.

The size-ratio, $R_f/R_i$, as defined in Section \ref{methods}, characterizes the length of the fingers with respect to the central, stable region.  As mentioned earlier, $R_f/R_i$ depends only on $\eta_{in}/\eta_{out}$ and is independent of $\lambda_c$ (\cite{irmgard, radar}). $R_f/R_i$ reaches a stable value as the pattern continues to grow; at long times, this ratio only increases only minimally. To make a consistent comparison between all pairs of fluids, $R_f/R_i$ is measured when $R_o = 8 cm$. 

\begin{figure}
\centering 
\begin{center} 
\includegraphics[width = 0.5\columnwidth]{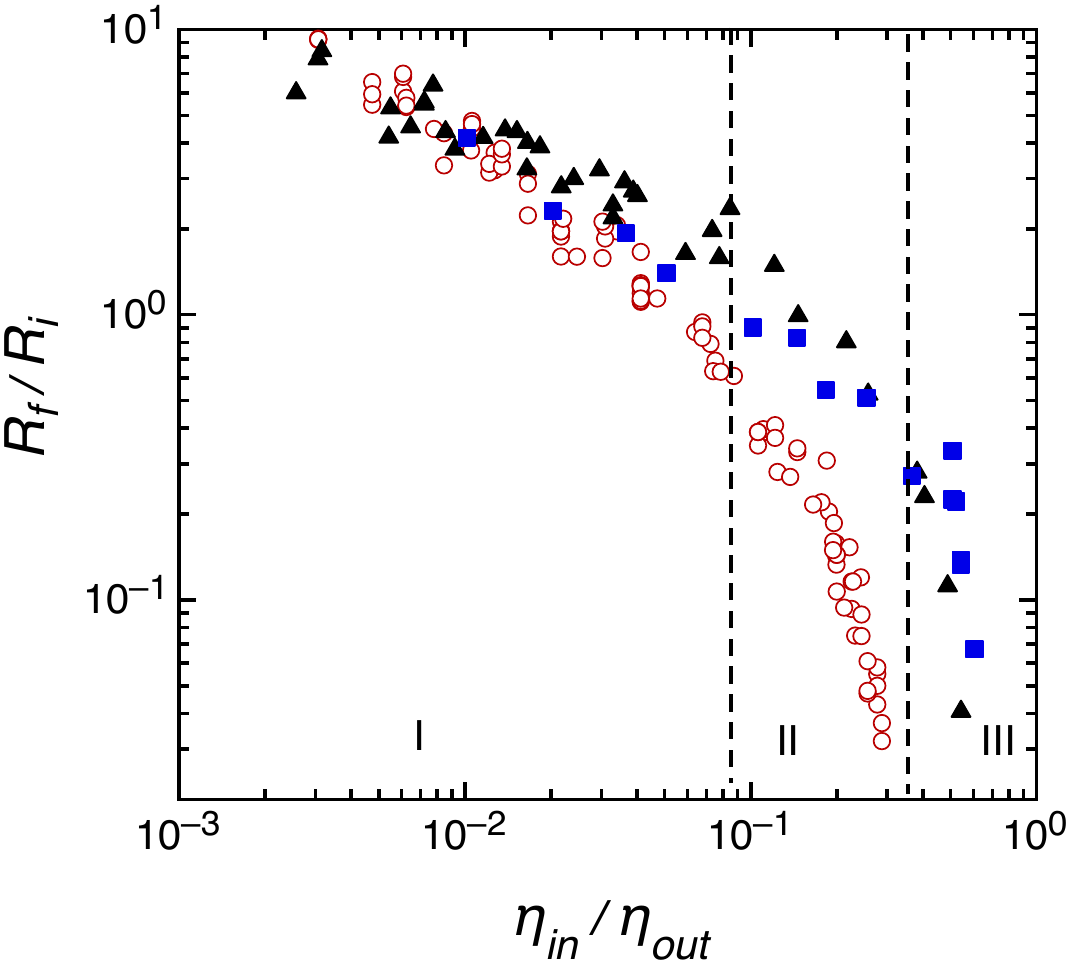} 
\caption{Dependence of the size-ratio, $R_f/R_i$, on the viscosity ratio, $\eta_{in}/\eta_{out}$, for three fluid systems with different interfacial tensions, $\sigma$. ($\blacktriangle$) - silicone oils displaced by glycerin-water (immiscible - high interfacial tension: $24 mN/m <\sigma< 29 mN/m$) (\textcolor {blue}{$\blacksquare$}) - silicone oils displaced by mineral oils (immiscible - low interfacial tension: $1 mN/m <\sigma< 1.2 mN/m$). (\textcolor {red}{$\circ$}) - glycerin-water mixtures (miscible - negligible interfacial tension). Dotted vertical lines show boundaries of different regimes in miscible displacements as identified by \cite{irmgard}. Immiscible displacements are similar irrespective of the different interfacial tensions; miscible fluids differs from immiscible ones at high viscosity ratios}
\label{data-all}
\end{center}
\end{figure}

Figure \ref{data-all} compares the dependence of the size-ratio, $R_f/R_i$, on the viscosity ratio, $\eta_{in}/\eta_{out}$, for three different fluid systems: two immiscible fluid pairs and one miscible pair. These three fluid systems demonstrate the effect of systematically lowering the interfacial tension, $\sigma$. Silicone oils/glycerin-water mixtures fluid pairs (immiscible), denoted by black triangles ($\blacktriangle$), have $24 mN/m < \sigma < 29 mN/m$. Their finger widths follow the prediction Eqn. \ref{saff-t}. Silicone oils/mineral oils fluid pairs (immiscible), the blue squares (\textcolor {blue}{$\blacksquare$}), have $1mN/m < \sigma < 1.2 mN/m$, such that the finger widths are governed by Eqn. \ref{patt}.  Miscible fluid data taken from \cite{irmgard}, is represented here using red open circles (\textcolor {red}{$\circ$}). These are mixtures of glycerol and water with different viscosities as both the inner and outer fluids.  They also have finger widths proportional to the plate spacing (Eqn. \ref{patt}). The figure also shows the boundary lines of the three miscible-fingering regimes previously identified \cite{irmgard}.

For low $\eta_{in}/\eta_{out}$ ($<$ 0.04), the three datasets in Fig. \ref{data-all} overlap. For $\eta_{in}/\eta_{out} > $ 0.04, the finger lengths found for the miscible fluids are smaller than those of immiscible fluids at the same viscosity ratio. However, there is no observable difference between the two sets of immiscible fluids with different interfacial tensions. As $\eta_{in}/\eta_{out} \rightarrow 1$, $R_f/R_i$ decreases towards zero for all the data sets but with the miscible fluids decreasing faster than the immiscible fluids.  

There are two critical differences between miscible and immiscible pairs. (i) When the interface shows a delayed onset of instability, this is followed by ``proportionate'' growth for miscible fluids. This is labeled regime II in the figure.  In this regime, the fingers grow slowly, in tandem with the stable central region. In contrast, for the immiscible fluids, after a delayed onset, the fingers grow faster than the central region. From Fig. \ref{data-all}, it can be seen that the region where the $R_f/R_i$ of the miscible and immiscible fluids begin to differ appreciably, corresponds approximately to the border between regimes I and II for miscible fluids. (ii) There are no data points for miscible fluids beyond $\eta_{in}/\eta_{out} = $ 0.33 and for immiscible fluids beyond $\eta_{in}/\eta_{out} = $ 0.6.  However, when larger Hele-Shaw plates are used, immiscible fluids with $\eta_{in}/\eta_{out} > $ 0.6, also become unstable, whereas for miscible fluids irrespective of the plate size no instability beyond $\eta_{in}/\eta_{out} = $ 0.33 is observed. Hence, the border for stability in miscible fluids is fixed by viscosity ratio alone, whereas it depends on the overall size of the plates in immiscible fluids. 

Thus, by using three pairs of liquids, I have systematically lowered the interfacial tension at the interface. This produces thinner and thinner fingers until the finger width is cut off by the gap spacing, $b$.  When the interfacial tension is negligible, as in the case of miscible fluids, there is a proportionate growth regime and a regime of complete stability.  By eliminating the stabilizing force or interfacial tension one has paradoxically produced more stable patterns.  One possible reason for this unusual stability, could lie in the fact that miscible fingering patterns show a gradation in color signifiying an incomplete displacement of the outer fluid as compared to immiscible fluids (\cite{lajeunesse19973d, irmgard}). 

\section{Cross-sectional shape of the fingers}
\label{tongue-section}

\begin{figure}
\centering 
\begin{center} 
\includegraphics[width = 0.9\columnwidth]{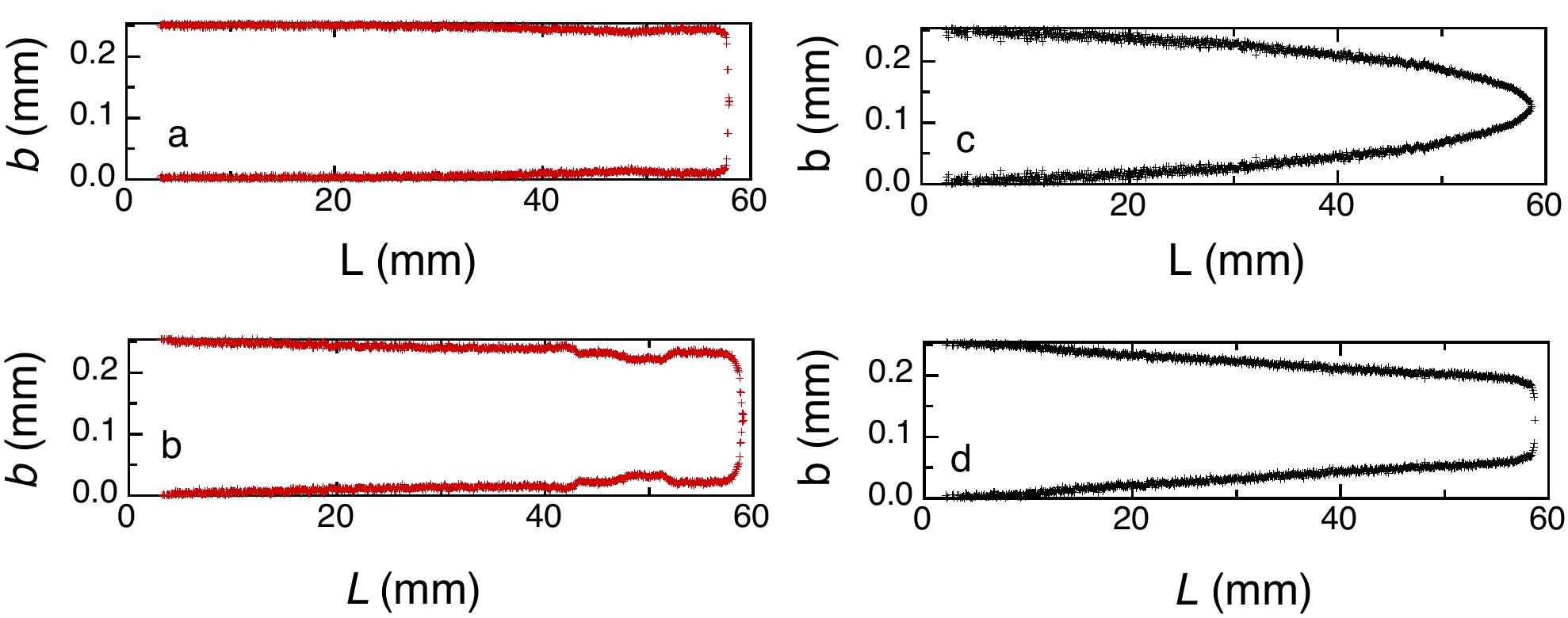} 
\caption{The profile of the displacing fluid in the direction perpendicular to the plates. (a), (b) show finger profiles for immiscible pairs of fluids: silicone oil displaced by mineral oil with a viscosity ratio (a) $\eta_{in}/\eta_{out}$ = 0.93 and (b) $\eta_{in}/\eta_{out}$ = 0.182. (c), (d) are displacement profiles taken from Ref. \cite{irmgard}. The viscosity ratios are: (c) $\eta_{in}/\eta_{out}$ = 1, (d) $\eta_{in}/\eta_{out}$ = 0.204. $L$ is the distance measured from the center of the plates. Immiscible displacements have a flat front at all viscosity ratios. In miscible displacements the shape of the front profile changes from rounded when the displacement is stable to flat when the displacement is unstable \cite{lajeunesse19973d}.}
\label{tongue}
\end{center}
\end{figure}

In this section, the cross-sectional profile of viscous fingers in miscible and immiscible fluids is compared. Figure \ref{tongue} shows the cross-sectional profile of viscous displacements when the interface is unstable (Fig. \ref{tongue}a,c) and when it is stable (in Fig. \ref{tongue}b,d). 

Looking at the cross-sectional profile of the tongue for immiscible fluids in Fig. \ref{tongue}a,b, we see not only that the displacement is almost complete, but also that the shape of the displacement front is flat. In Fig. \ref{tongue}a the tongue fills more than 97\% of the plate spacing and in Fig. \ref{tongue}b the width of the tongue is 83\% of the gap-width at its tip and about 74\% of the gap-width at it's narrowest point. The capillary number, $Ca \equiv \Delta \eta U/ \sigma$, is approximately 0.001 and 0.8 for Fig. \ref{tongue}a and b respectively. These values for the widths of the tongue are in the same range as the measurements by \cite{FLM:392852} for immiscible displacements at high capillary numbers. There is also no marked change in the shape or size of the displacing front as we transition from fingering to delayed fingering to stable pattern. 

In contrast, miscible fluids produce a less complete displacement of the outer fluid, as shown in Fig. \ref{tongue}c,d. Notably, in Fig. \ref{tongue}c, where there is a stable interface for miscible fluids, the thickness of the tongue smoothly reduces to zero at its tip; at the onset of fingering the tip of this profile becomes flat \cite{irmgard, lajeunesse19973d}. 

This could explain why immiscible fluids do not show completely stable patterns like miscible fluids. The apparent stability of immiscible fluids is caused because the fingering onset has been sufficiently delayed, such that it does not become unstable within the boundary of the experimental plates. Hence  it does not correspond to a new regime, as in the case of miscible fluids. However, the tongue profile does not explain the delay in the onset of instability that is found in both miscible and immiscible fluids. 

\section{Onset of instability in immiscible displacements}
\label{onset-para}

\begin{figure}
\centering 
\begin{center} 
\includegraphics[width = 0.65\columnwidth]{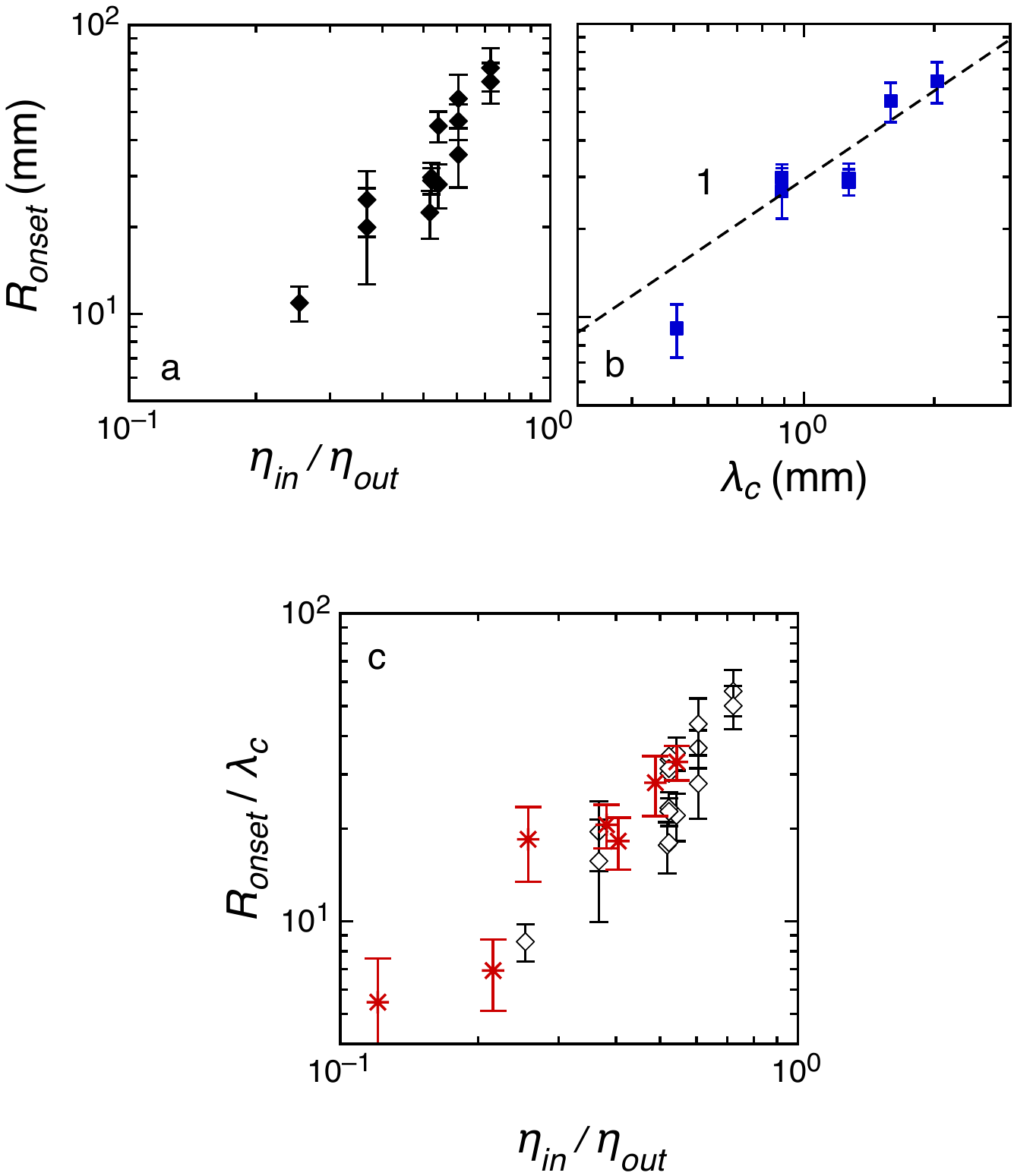} 
\caption{ Dependence of the onset radius, $R_{onset}$, in immiscible fluids on the viscosity ratio, $\eta_{in}/\eta_{out}$ and the most unstable wavelength, $\lambda_c$. (a) shows the variation of $R_{onset}$ on $\eta_{in}/\eta_{out}$ for pairs of silicone-oil/mineral-oil fluids at $\lambda_c = $ 1.2 $\pm$0.1 mm, (b) shows the dependence of $R_{onset}$ on $\lambda_c$ when $\eta_{in}/\eta_{out} =$ 0.52 for a silicone-oil/mineral-oil pair. $R_{onset}$ has an approximately linear dependence on $\lambda_c$ (c) shows $R_{onset}/\lambda_c$ as a function of $\eta_{in}/\eta_{out}$ for two sets of immiscible fluids: ($\diamond$) - silicone oils/mineral oils with $1 mN/m <\sigma< 1.2 mN/m$ and (\textcolor {red}{$\ast$}) - silicone oils/glycerin-water mixtures with $24 mN/m <\sigma< 29 mN/m$. $R_{onset}/\lambda_c$ has a power-law dependence with an exponent 1.5$\pm $ 0.2}
\label{Ronset}
\end{center}
\end{figure}

Delayed onset in miscible fluids has been shown to depend on the viscosity ratio, $\eta_{in}/\eta_{out}$ \cite{irmgard}. Here, an analysis of the onset radius, $R_{onset}$ in immiscible fluids is discussed.  Figure \ref{Ronset}, shows the dependence of $R_{onset}$ on both the viscosity ratio, $\eta_{in}/\eta_{out}$, and the most unstable wavelength, $\lambda_c$.  In Fig. \ref{Ronset}a, $\eta_{in}/\eta_{out}$ is varied while keeping $\lambda_c$ constant. This is done by using pairs of silicone-oil/mineral-oil fluids where $\lambda_c $ is set by the plate spacing.   $R_{onset}$ increases rapidly with $\eta_{in}/\eta_{out}$. 

In order to vary $\lambda_c$ independently of $\eta_{in}/\eta_{out}$, the same silicone-oil / mineral-oil pair of fluids at $\eta_{in}/\eta_{out} = $ 0.52 is used and the plate spacing, $b$ is varied. Figure \ref{Ronset}b shows that $R_{onset}$ varies approximately linearly with $\lambda_c$. This allows a rescaling of the data for all viscosity ratios using $\lambda_c$, as shown in Fig. \ref{Ronset}c. In this figure, data is also included with the higher interfacial tension pairs of fluids (silicone oils/glycerol-water mixtures) marked by the red crosses (\textcolor {red}{$\ast$}). The open diamonds ($\diamond$) are data taken with silicone oils/mineral oils. The dimensionless parameter $R_{onset}/\lambda_c$ increases with $\eta_{in}/\eta_{out}$:
\begin{equation}
R_{onset}/\lambda_c \propto (\eta_{in}/\eta_{out})^{\gamma}
\end{equation}
where $\gamma$ = 1.5$\pm$0.2

Although there are no data points for immiscible fluids in Figs. \ref{data-all} and \ref{Ronset} (which were obtained using 14$cm$ radius plates) at $\eta_{in}/\eta_{out} = $ 1, I  do find unstable patterns at higher viscosity ratios by using larger plates, such that the radius of the plates remains larger than $R_{onset}$. Hence, the Saffman-Taylor stability criterion of viscous instability for viscosity ratios smaller than 1, is still valid for immiscible fluids.  

Delay in the onset of the instability, has been previously observed and attributed to the effect of radial geometry \cite{al2012control}. The values of $R_{onset}$ reported here for immiscible fluids vary between 1 cm to 9 cm and are comparable to those reported for miscible fluids \cite{irmgard}.  These are one to two orders of magnitude larger than the previously suggested effects due to geometry, which we calculate to be no greater than 1 mm for all our experiments.  Thus the delay in onset of instability is a characteristic feature of the viscous-fingering instability at high $\eta_{in}/\eta_{out}$ for both miscible and immiscible fluids.  Understanding the physical processes behind this phenomenon may provide a deeper insight into this classic problem.

\section{Conclusion}
Viscous fingering has been studied as a prototypical system for pattern formation since the work of Saffman and Taylor (\cite{saffman1958penetration, nittman1985fractal, daccord1986radial, bensimon1986viscous, PhysRevA.33.1302, PhysRevLett.63.984, casademunt2004viscous, hinrichsen1989self}). While there is a large literature dedicated to understanding the most unstable wavelength, $\lambda_c$ and its contribution to pattern formation (\cite{chen1987radial, homsy1987viscous, paterson1981radial, tabeling1987experimental, mccloud1995experimental, miranda1998radial, goyal2006miscible, moore2002fluctuations}), less attention has been paid to the regime where the interface begins to be unstable at viscosity ratios close to one. 

Recently, a wide range of viscosity ratios, spanning roughly three orders of magnitude, were studied by \cite{irmgard, radar}.  This revealed that the viscosity ratio, $\eta_{in}/\eta_{out}$, controls the length of fingering pattern with respect to a central stable pattern in both miscible and immiscible fluids. This paper has focused on the differences between the miscible and immiscible cases, especially as $\eta_{in}/\eta_{out} \rightarrow 1$. The role played by interfacial tension, is studied by looking at three sets of fluids - (i) immiscible with high interfacial tension, (ii) immiscible with low interfacial tension and (iii) miscible fluids with negligible interfacial tension. For immiscible fluids with high interfacial tension, the unstable wavelength follows Saffman-Taylor equation (Eqn. \ref{saff-t}. For the other two cases, the most unstable wavelength gets cut-off by the plate spacing. 

Interfacial tension plays no role in determining the length of fingers in all three scenarios. It does however prevent the formation of a 3D thin tongue profile found in miscible fluids. Hence, while miscible fluids have a completely stable regime for all $\eta_{in}/\eta_{out} > $ 0.33, no such regime change occurs in immiscible fluids.

A pronounced delay in the onset of fingering is found in all three cases. However, in immiscible fluids once the fingers do form after the delay, they grow very rapidly.  This is in contrast to the slow ``proportionate'' growth regime found in miscible fluids. The delayed onset radius of fingering in immiscible fluids is much larger than what has been predicted by previous theoretical work (\cite{al2012control}). This onset radius is proportional to the most unstable wavelength and increases approximately as a power law with the viscosity ratio, $\eta_{in}/\eta_{out}$. 

Varying $\eta_{in}/\eta_{out}$ has provided a useful way to study the transition between unstable fingering patterns and the stable regime of complete fluid displacement. It reveals a new perspective to the half-century old Saffman-Taylor problem and has left several mysteries yet to be answered. Understanding what suppresses the onset of fingering at high $\eta_{in}/\eta_{out}$ and the mechanism by which miscible fluids successfully suppress fingering, as compared to immiscible fluids, could give insight into how the viscous fingering instability can be suppressed in industrial processes, where it can produce undesired results. 

\section{Acknowledgments}
I would like to sincerely thank Sidney Nagel for his guidance. I especially thank Irmgard Bischofberger for her support in this project. I would also like to thank Justin Burton, Julian Freed-Brown, Wendy Zhang, Todd Dupont, Leo Kadanoff, Paul Wiegmann and Tom Witten for helpful discussions. This work was supported by NSF Grant DMR-1404841 and by NSF MRSEC DMR-1420709.




\bibliographystyle{elsarticle-harv} 
\bibliography{Radha-a-comparison-PD-els-final}

\begin{thebibliography}{43}
\expandafter\ifx\csname natexlab\endcsname\relax\def\natexlab#1{#1}\fi
\expandafter\ifx\csname url\endcsname\relax
  \def\url#1{\texttt{#1}}\fi
\expandafter\ifx\csname urlprefix\endcsname\relax\def\urlprefix{URL }\fi

\bibitem[{Al-Housseiny et~al.(2012)Al-Housseiny, Tsai, and
  Stone}]{al2012control}
Al-Housseiny, T.~T., Tsai, P.~A., Stone, H.~A., 2012. Control of interfacial
  instabilities using flow geometry. Nat. Phys. 8~(10), 747--750.

\bibitem[{Araktingi et~al.(1993)Araktingi, Orr~Jr,
  et~al.}]{araktingi1993viscous}
Araktingi, U.~G., Orr~Jr, F., et~al., 1993. Viscous fingering in heterogeneous
  porous media. SPE Advanced Technology Series 1~(01), 71--80.

\bibitem[{Arashiro and Demarquette(1999)}]{ARASHIRO1999}
Arashiro, E.~Y., Demarquette, N.~R., 01 1999. {Use of the pendant drop method
  to measure interfacial tension between molten polymers}. {Mater. Res.} 2, 23
  -- 32.

\bibitem[{Arn\'eodo et~al.(1989)Arn\'eodo, Couder, Grasseau, Hakim, and
  Rabaud}]{PhysRevLett.63.984}
Arn\'eodo, A., Couder, Y., Grasseau, G., Hakim, V., Rabaud, M., Aug 1989.
  Uncovering the analytical saffman-taylor finger in unstable viscous fingering
  and diffusion-limited aggregation. Phys. Rev. Lett. 63, 984--987.
\newline\urlprefix\url{https://link.aps.org/doi/10.1103/PhysRevLett.63.984}

\bibitem[{Bensimon(1986)}]{PhysRevA.33.1302}
Bensimon, D., Feb 1986. Stability of viscous fingering. Phys. Rev. A 33,
  1302--1308.
\newline\urlprefix\url{https://link.aps.org/doi/10.1103/PhysRevA.33.1302}

\bibitem[{Bensimon et~al.(1986)Bensimon, Kadanoff, Liang, Shraiman, and
  Tang}]{bensimon1986viscous}
Bensimon, D., Kadanoff, L.~P., Liang, S., Shraiman, B.~I., Tang, C., 1986.
  Viscous flows in two dimensions. Rev. Mod. Phys. 58~(4), 977.

\bibitem[{Bischofberger et~al.(2014)Bischofberger, Ramachandran, and
  Nagel}]{irmgard}
Bischofberger, I., Ramachandran, R., Nagel, S.~R., 2014. Fingering versus
  stability in the limit of zero interfacial tension. Nat. Comm. 5, 5265.

\bibitem[{Bischofberger et~al.(2015)Bischofberger, Ramachandran, and
  Nagel}]{radar}
Bischofberger, I., Ramachandran, R., Nagel, S.~R., 2015. An island of stability
  in a sea of fingers: Emergent large-scale features of the viscous flow
  instability. Soft Matter.

\bibitem[{Casademunt(2004)}]{casademunt2004viscous}
Casademunt, J., 2004. Viscous fingering as a paradigm of interfacial pattern
  formation: Recent results and new challenges. Chaos: An Interdisciplinary
  Journal of Nonlinear Science 14~(3), 809--824.

\bibitem[{Chen(1987)}]{chen1987radial}
Chen, J.-D., 1987. Radial viscous fingering patterns in hele-shaw cells. Exp.
  Fluids. 5~(6), 363--371.

\bibitem[{Cheng et~al.(2008)Cheng, Xu, Patterson, Jaeger, and
  Nagel}]{cheng2008towards}
Cheng, X., Xu, L., Patterson, A., Jaeger, H.~M., Nagel, S.~R., 2008. Towards
  the zero-surface-tension limit in granular fingering instability. Nat. Phys.
  4~(3), 234--237.

\bibitem[{Cinar et~al.(2007)Cinar, Riaz, Tchelepi,
  et~al.}]{cinar2007experimental}
Cinar, Y., Riaz, A., Tchelepi, H.~A., et~al., 2007. Experimental study of co2
  injection into saline formations. In: SPE Annual Technical Conference and
  Exhibition. Society of Petroleum Engineers.

\bibitem[{Daccord et~al.(1986)Daccord, Nittmann, and
  Stanley}]{daccord1986radial}
Daccord, G., Nittmann, J., Stanley, H.~E., 1986. Radial viscous fingers and
  diffusion-limited aggregation: Fractal dimension and growth sites. Phys. Rev.
  Lett. 56~(4), 336.

\bibitem[{Dhar and Sadhu(2013)}]{dhar2013sandpile}
Dhar, D., Sadhu, T., 2013. A sandpile model for proportionate growth. J. Stat.
  Mech 2013~(11), P11006.

\bibitem[{{\=E}rglis et~al.(2013){\=E}rglis, Tatulcenkov, Kitenbergs,
  Petrichenko, Ergin, Watz, and C{\=e}bers}]{erglis2013magnetic}
{\=E}rglis, K., Tatulcenkov, A., Kitenbergs, G., Petrichenko, O., Ergin, F.,
  Watz, B., C{\=e}bers, A., 2013. Magnetic field driven micro-convection in the
  hele-shaw cell. Journal of Fluid Mechanics 714, 612--633.

\bibitem[{Ferer et~al.(2004)Ferer, Ji, Bromhal, Cook, Ahmadi, and
  Smith}]{PhysRevE.70.016303}
Ferer, M., Ji, C., Bromhal, G.~S., Cook, J., Ahmadi, G., Smith, D.~H., Jul
  2004. Crossover from capillary fingering to viscous fingering for immiscible
  unstable flow: Experiment and modeling. Phys. Rev. E 70, 016303.
\newline\urlprefix\url{https://link.aps.org/doi/10.1103/PhysRevE.70.016303}

\bibitem[{Goyal and Meiburg(2006)}]{goyal2006miscible}
Goyal, N., Meiburg, E., 2006. Miscible displacements in hele-shaw cells:
  two-dimensional base states and their linear stability. J. Fluid. Mech. 558,
  329--355.

\bibitem[{Hill et~al.(1952)}]{hill1952channeling}
Hill, S., et~al., 1952. Channeling in packed columns. Chem. Eng. Sci. 1~(6),
  247--253.

\bibitem[{Hinrichsen et~al.(1989)Hinrichsen, Maloy, Feder, and
  Jossang}]{hinrichsen1989self}
Hinrichsen, E.~L., Maloy, K., Feder, J., Jossang, T., 1989. Self-similarity and
  structure of dla and viscous fingering clusters. Journal of Physics A:
  Mathematical and General 22~(7), L271.

\bibitem[{Homsy(1987)}]{homsy1987viscous}
Homsy, G.~M., 1987. Viscous fingering in porous media. Annu. Rev. Fluid. Mech.
  19~(1), 271--311.

\bibitem[{Kitenbergs et~al.(2015)Kitenbergs, Tatulcenkovs, {\=E}rglis,
  Petrichenko, Perzynski, and C{\=e}bers}]{kitenbergs2015magnetic}
Kitenbergs, G., Tatulcenkovs, A., {\=E}rglis, K., Petrichenko, O., Perzynski,
  R., C{\=e}bers, A., 2015. Magnetic field driven micro-convection in the
  hele-shaw cell: the brinkman model and its comparison with experiment.
  Journal of Fluid Mechanics 774, 170--191.

\bibitem[{Lajeunesse et~al.(1997)Lajeunesse, Martin, Rakotomalala, and
  Salin}]{lajeunesse19973d}
Lajeunesse, E., Martin, J., Rakotomalala, N., Salin, D., 1997. 3d instability
  of miscible displacements in a hele-shaw cell. Phys. Rev. Lett. 79~(26),
  5254.

\bibitem[{Li et~al.(2009)Li, Lowengrub, Fontana, and
  Palffy-Muhoray}]{li2009control}
Li, S., Lowengrub, J.~S., Fontana, J., Palffy-Muhoray, P., 2009. Control of
  viscous fingering patterns in a radial hele-shaw cell. Phys. Rev. Lett.
  102~(17), 174501.

\bibitem[{Lindner et~al.(2002)Lindner, Bonn, Poir{\'e}, Amar, and
  Meunier}]{lindner2002viscous}
Lindner, A., Bonn, D., Poir{\'e}, E.~C., Amar, M.~B., Meunier, J., 2002.
  Viscous fingering in non-newtonian fluids. Journal of Fluid Mechanics 469,
  237--256.

\bibitem[{McCloud and Maher(1995)}]{mccloud1995experimental}
McCloud, K.~V., Maher, J.~V., 1995. Experimental perturbations to
  saffman-taylor flow. Phys. Rep. 260~(3), 139--185.

\bibitem[{Miranda and Widom(1998)}]{miranda1998radial}
Miranda, J., Widom, M., 1998. Radial fingering in a hele-shaw cell: a weakly
  nonlinear analysis. Physica D 120~(3), 315--328.

\bibitem[{Moore et~al.(2002)Moore, Juel, Burgess, McCormick, and
  Swinney}]{moore2002fluctuations}
Moore, M.~G., Juel, A., Burgess, J.~M., McCormick, W., Swinney, H.~L., 2002.
  Fluctuations in viscous fingering. Phys. Rev. E 65~(3), 030601.

\bibitem[{Nittman et~al.(1985)Nittman, Daccord, and
  Stanley}]{nittman1985fractal}
Nittman, J., Daccord, G., Stanley, M., 1985. Fractal growth of viscous fingers:
  quantitative characterization of a fluid instability phenomenon. Nature 314,
  391.

\bibitem[{Orr and Taber(1984)}]{orr1984use}
Orr, F., Taber, J., 1984. Use of carbon dioxide in enhanced oil recovery.
  Science 224~(4649), 563--569.

\bibitem[{Paterson(1981)}]{paterson1981radial}
Paterson, L., 1981. Radial fingering in a hele shaw cell. J. Fluid. Mech. 113,
  513--529.

\bibitem[{Paterson(1985)}]{paterson1985fingering}
Paterson, L., 1985. Fingering with miscible fluids in a hele shaw cell. Phys.
  Fluids. 28~(1), 26--30.

\bibitem[{Pihler-Puzovi{\'c} et~al.(2012)Pihler-Puzovi{\'c}, Illien, Heil, and
  Juel}]{pihler2012suppression}
Pihler-Puzovi{\'c}, D., Illien, P., Heil, M., Juel, A., 2012. Suppression of
  complex fingerlike patterns at the interface between air and a viscous fluid
  by elastic membranes. Physical review letters 108~(7), 074502.

\bibitem[{Praud and Swinney(2005)}]{praud2005fractal}
Praud, O., Swinney, H.~L., 2005. Fractal dimension and unscreened angles
  measured for radial viscous fingering. Phys. Rev. E 72~(1), 011406.

\bibitem[{Sadhu and Dhar(2012)}]{sadhu2012modelling}
Sadhu, T., Dhar, D., 2012. Modelling proportionate growth. Curr. Sci. 103~(5).

\bibitem[{Saffman and Taylor(1958)}]{saffman1958penetration}
Saffman, P.~G., Taylor, G., 1958. The penetration of a fluid into a porous
  medium or hele-shaw cell containing a more viscous liquid. P. Roy. Soc.
  A-Math. Phy. 245~(1242), 312--329.

\bibitem[{Setu et~al.(2013)Setu, Zacharoudiou, Davies, Bartolo, Moulinet,
  Louis, Yeomans, and Aarts}]{setu2013viscous}
Setu, S.~A., Zacharoudiou, I., Davies, G.~J., Bartolo, D., Moulinet, S., Louis,
  A.~A., Yeomans, J.~M., Aarts, D.~G., 2013. Viscous fingering at ultralow
  interfacial tension. Soft Matter 9~(44), 10599--10605.

\bibitem[{Tabeling et~al.(1987{\natexlab{a}})Tabeling, Zocchi, and
  Libchaber}]{tabeling1987experimental}
Tabeling, P., Zocchi, G., Libchaber, A., 1987{\natexlab{a}}. An experimental
  study of the saffman-taylor instability. J. Fluid. Mech. 177, 67--82.

\bibitem[{Tabeling et~al.(1987{\natexlab{b}})Tabeling, Zocchi, and
  Libchaber}]{FLM:392852}
Tabeling, P., Zocchi, G., Libchaber, A., 4 1987{\natexlab{b}}. An experimental
  study of the saffman-taylor instability. J. Fluid Mech. 177, 67--82.

\bibitem[{Than et~al.(1988)Than, Preziosi, Josephl, and
  Arney}]{than1988measurement}
Than, P., Preziosi, L., Josephl, D., Arney, M., 1988. Measurement of
  interfacial tension between immiscible liquids with the spinning road
  tensiometer. J. Colloid. Interf. Sci. 124~(2), 552--559.

\bibitem[{Wang et~al.(2004)Wang, Jury, Tuli, and Kim}]{wang2004unstable}
Wang, Z., Jury, W.~A., Tuli, A., Kim, D.-J., 2004. Unstable flow during
  redistribution. Vadose Zone J. 3~(2), 549--559.

\bibitem[{Weitz et~al.(1987)Weitz, Stokes, Ball, and
  Kushnick}]{weitz1987dynamic}
Weitz, D., Stokes, J., Ball, R., Kushnick, A., 1987. Dynamic capillary pressure
  in porous media: Origin of the viscous-fingering length scale. Physical
  review letters 59~(26), 2967.

\bibitem[{White and Ward(2012)}]{white2012co2}
White, A.~R., Ward, T., 2012. Co2 sequestration in a radial hele-shaw cell via
  an interfacial chemical reaction. Chaos 22~(3), 037114.

\bibitem[{Yang and Yortsos(1997)}]{yang1997asymptotic}
Yang, Z., Yortsos, Y.~C., 1997. Asymptotic solutions of miscible displacements
  in geometries of large aspect ratio. Physics of Fluids 9~(2), 286--298.

\end{thebibliography}




%
\end{document}